\RequirePackage{fix-cm}
\documentclass[smallextended]{svjour3}       
\smartqed  
\usepackage{graphicx}
\usepackage{enumerate}
\usepackage{float}
\usepackage{caption}
\usepackage[justification=centering]{caption}
\usepackage{amsmath}
\usepackage{comment}
\usepackage{mathrsfs}
\usepackage{color}
\usepackage[colorlinks,urlcolor=blue,citecolor=blue]{hyperref}
\usepackage{cite}
\usepackage{multirow}
\usepackage{colortbl}
\definecolor{mygray}{gray}{.70}

\begin{document}

\title{Verifiable Threshold Quantum Secret Sharing with Sequential Communication
}



\author{Changbin Lu$^1$          \and
        Fuyou Miao$^1$          \and
        Junpeng Hou$^2$          \and
        Keju Meng$^1$
}

\institute{Changbin Lu \at
              \email{lcb@mail.ustc.edu.cn}           
           \and
           Fuyou Miao \at
              \email{mfy@ustc.edu.cn}
           \and
           Junpeng Hou \at
             \email{Junpeng.Hou@utdallas.edu}
           \and
           Keju Meng \at
                \email{mkj@mail.ustc.edu.cn}
           \and
           $^1$ School of Computer Science and Technology, University of Science and Technology of China, Hefei, China\\
          $^2$ Department of Physics, The University of Texas at Dallas, Richardson, Texas 75080-3021, USA
}
\date{Received: date / Accepted: date}

\maketitle

\begin{abstract}
A ($t$, $n$) threshold quantum secret sharing (QSS) is proposed based on a single $d$-level quantum system. It enables the ($t$, $n$) threshold structure based on Shamir's secret sharing and simply requires sequential communication in $d$-level quantum system to recover secret. Besides, the scheme provides a verification mechanism which employs an additional qudit to detect cheats and eavesdropping during secret reconstruction, and allows a participant to use the share repeatedly. Analyses show that the proposed scheme is resistant to typical attacks. Moreover, the scheme is scalable in participant number and easier to realize compared to related schemes. More generally, our scheme also presents a generic method to construct new ($t$, $n$) threshold QSS schemes based on $d$-level quantum system from other classical threshold secret sharing.
\end{abstract}

\section{Introduction}
Suppose a dealer needs to share a secret message among a group of users but does not want any single user to have the whole secret. How can the dealer achieve this goal without directly allocating a copy to any user? A desirable method is to distribute a shadow derived from the secret to each user, such that some certain number of users can cooperate to recover the  secret while fewer users cannot obtain any information of the secret. To address the problem of confidentiality and robustness in keeping a secret among users, Shamir \cite{Shamir1979} and Blakely \cite{Blakley1979} proposed the well-defined ($t$, $n$) threshold secret sharing [($t$, $n$)-SS] scheme independently in 1979, with two criteria should be respected. The first one is reliability, meaning the scheme should allow at least $t$ users can recover the secret. The second criterion is confidentiality, it means less than $t$ users should not gain any information about the secret (even with unlimited computation resource). Today, ($t$, $n$)-SS has become a fundamental cryptographic primitive and been widely used in many applications such as group authentication \cite{Harn2013}, threshold signature \cite{Harn1994,Boldyreva2002}, group key agreement \cite{Liu2016}, threshold encryption \cite{Desmedt1994}, secure multiparty computation \cite{Patel2016}, etc.

In recent years, quantum cryptography has attracted much attention due to its inherent security. Based on physical laws such as Heisenberg uncertainty principle and the resulted quantum no-cloning theorem, quantum cryptographic protocols are able to provide unconditional security while classical ones usually have computational security based on computational complexity. Thus, using quantum-information-assisted schemes, $i.e.$, quantum secret sharing (QSS), to share secrets among users is more reliable and promising. Furthermore, QSS provides a robust and secure solution for quantum state storage and computation \cite{Cleve1999}. Such scheme was first proposed by Hillery $et$ $al$. \cite{Hillery1999} in 1999, which takes advantage of a three-qubit entangled Greenberger-Horne-Zeilinger (GHZ) state. In the scheme, a GHZ triplet is split and each of the other two participants get a particle. Both participants are allowed to measure their particles in either $x$ or $y$ basis (natural basis) and their results are combined to give the dealer's measurement result. In this way, a joint secret is established between the dealer and corresponding users. Following the similar idea, the QSS is further generalized to $d$-level platform \cite{Yu2008} by utilizing multiparticle ($>3$) entanglement GHZ state. Subsequently, another QSS \cite{Bai2017} was proposed using the $d$-dimensional GHZ state in a different way, in which participants use an X-basis measurement and classical communication to distinguish two orthogonal states and reconstruct the original secret. However, these entanglement-based schemes are all poor in scalability because it gets difficult to keep quantum correlations among more and more participants. Obviously, supporting such QSS with more participants requires more entangled state to be prepared, however, with the increase of the number of qubits, entangled state preparation becomes much more difficult. Moreover, quantum correlations are prone to be spoiled through interacting with environment.

Currently, many existing QSS schemes are of ($n$, $n$) type \cite{Hillery1999,Guo2003,Yu2008,Bai2017,Tavakoli2015,Hsu2003,Markham2008,Karimipour2015,Yang2013,Lai2016_2,Kogias2017} including some experimental
demonstrations \cite{Tittel2001,Schmid2005,Chen2005,Lu2016}, which requires all $n$ shareholders, instead of any $t$ or more than $t$ shareholders, to cooperate in recovering the secret. Therefore, they are less flexible than ($t$, $n$) ones and has limited applications. Since the first threshold QSS \cite{Cleve1999} was proposed, there have been mainly two methods to construct threshold QSS. The first method is purely using some special quantum systems \cite{Cleve1999,Lance2003,Lau2013,Wu2014,Rahaman2015,Yang2015,Wang2017} in schemes' construction. For example, in the seminal work \cite{Cleve1999}, an arbitrary three-dimensional quantum state (or qutrit) was employed to construct a (2, 3) threshold scheme, which  maps the private qutrit to three qutrits and each resulted qutrit is taken as a share. The second method is incorporating classical threshold SS with quantum operations and thus keep ($t$, $n$) threshold structure \cite{Tokunaga2005,Qin2015,Song2017,Lu2018,Lai2016}. These schemes employ quantum operations to embed private value and shares of classical threshold SS into quantum states, such that  $t$ or more than $t$ participants can recover the initial quantum state to gain the secret only after they each complete their operations. The first threshold QSS scheme based on Shamir's ($t$, $n$)-SS was proposed in \cite{Tokunaga2005}. In this scheme, a secret is initially embedded into quantum states; then, any $t$ or more than $t$ participants sequentially apply Hadamard transformation and proper rotation operations on the quantum state; finally  the secret can be regained after applying certain measurements on the state. Several quantum computation algorithms, such as phase shift operation or Quantum Fourier Transform, are also introduced to embed classical shares into quantum states \cite{Qin2015,Song2017,Lu2018}.

In this paper, we propose a ($t$, $n$) threshold QSS scheme based on a single $d$-level quantum system, where dimension $d$ is an odd prime number. In our scheme,  the dealer generates $n$ shares from a secret and allocates each share to shareholders as in Shamir's ($t$, $n$)-SS. To recover the secret, at least $t$ participants perform proper unitary operations (in some order) sequentially on a vector of a set of Mutually Unbiased (orthonormal) Bases (MUBs); subsequently, the qudit is measured in an appointed basis by the last participant. After the announcement of measurement result, participants exchange random numbers embedded in the qudit such that they can recover the dealer's secret. To guarantee the security against eavesdropping and cheats, a verification mechanism is established which uses an additional qudit to check the consistency of recovered secrets. Compared with existing QSS schemes, our scheme stands out for the following properties:
\begin{enumerate}[($i$)]
 \item It is more general and practicable than $2$-level QSS; moreover, private shares can be used repeatedly.

 \item It is scalable in the number of participants compared with schemes based on entangled states.

 \item As a ($t$, $n$) threshold QSS, it is more flexible in application than ($n$, $n$)-QSS;

 \item Based on the verification mechanism, it does not depend on any trusted third party and is able to detect any cheat and eavesdropping during secret reconstruction.

 \item Other classical ($t$, $n$)-SS schemes can be used to replace Shamir's scheme while keeping all the aforementioned properties.
\end{enumerate}
\section{Secret sharing based on a single $d$-level quantum system}

\subsection{The cyclic property of the MUBs}
In this paper, we construct a $(t,n)$ threshold quantum secret sharing scheme based on a set of MUBs which has the cyclic property \cite{Tavakoli2015}.  It has been proven that $d+1$ MUBs can be found in a $d$-dimension complex vector space if $d$ is an odd prime \cite{Ivanovic1981,Wootters1989}. Besides the computational basis $\{ \left| j \right\rangle,j = 0,1,...,d - 1\} $, the explicit forms of the remaining $d$ sets of MUBs are $\left| {{\varphi _l}^k} \right\rangle  = \frac{1}{{\sqrt d }}\sum_{j = 0}^{d - 1} {{\omega ^{j(l + kj)}}\left| j \right\rangle } $, where $k = 0,1,...,d - 1$ labels the basis, $l = 0,1,...,d - 1$ enumerates the vectors of the given basis and $\omega=e^{2\pi i/d}$ is the $d$th root of unity. For any two values $l \ne l'$, these MUBs have the following property
\begin{equation}
\left\langle {{\varphi _l}^k} \right|\left. {{\varphi _{l'}}^k} \right\rangle=0,
\end{equation}
and this indicates that ${ \left| \varphi_l^k \right\rangle} $ may serve as a basis. Moreover, they are mutually unbiased because
\begin{equation} \label{2}
{\left| {\left\langle {{\varphi _l}^k} \right|\left. {{\varphi _{l'}}^{k'}} \right\rangle } \right|^2}=\frac{1}{d}
\end{equation}
holds for $l \ne l'$ and $k \ne k'$. Beside from the viewpoint of bra-ket notation, Eq. (\ref{2}) can also be inferred from Number Theory due to $\left| {\sum\nolimits_{j = 0}^{d - 1} {{\omega ^{pj + q{j^2}}}} } \right| = \sqrt d$ for $p,q \in Z$, $q \ne 0$ and prime number $d$.

The set of MUBs has a cyclic property, $i.e.$, there exist unitary operations $U_{l'k'}$ for any $l',k' \in \{ 0,1,...,d - 1\} $ transforming a given vector $\left| {{\varphi _l}^k} \right\rangle $ into $\left| {{\varphi _{l + l'}}^{k + k'}} \right\rangle $. Specifically, the operations ${X_d} = \sum\nolimits_{r = 0}^{d - 1} {{\omega ^r}\left| r \right\rangle } \left\langle r \right|$ and ${Y_d} = \sum\nolimits_{r = 0}^{d - 1} {{\omega ^{{r^2}}}\left| r \right\rangle } \left\langle r \right|$ can transform the vector $\left| {{\varphi _l}^k} \right\rangle $ into $\left| {{\varphi _{l + 1}}^k} \right\rangle $ and $\left| {{\varphi _l}^{k + 1}} \right\rangle $ respectively due to
\begin{equation}
\begin{aligned}
{X_d}\left| {{\varphi _l}^k} \right\rangle
&=\frac{1}{{\sqrt d }}\sum\limits_{r = 0}^{d - 1} {{\omega ^r}\left| r \right\rangle \left\langle r \right|} \sum\limits_{j = 0}^{d - 1} {{\omega ^{j(l + kj)}}} \left| j \right\rangle  \\
&=\frac{1}{{\sqrt d }}\sum\limits_{r,j = 0}^{d - 1} {{\omega ^r}{\omega ^{j(l + kj)}}\left| r \right\rangle } {\delta _{rj}} \\
&=\frac{1}{{\sqrt d }}\sum\limits_{j = 0}^{d - 1} {{\omega ^{j[(l + 1) + kj]}}\left| j \right\rangle }  = \left| {{\varphi _{l + 1}}^k} \right\rangle ,
\end{aligned}
\end{equation}
while the correctness of $Y_d$ can be proven in the same way. As a result, the unitary operations $U{}_{l'k'}$ is just the combination of those two operators ${U_{l'k'}} = {X_d}^{l'}{Y_d}^{k'}$ and note that $[X_d,Y_d]=0$ guarantees the definition of exponents.
\subsection{Threshold secret sharing based on qudits}
\subsubsection{Overview}

The scheme is constructed based on the cyclic property of MUBs and classical ($t, n$)-SS.
Initially, each shareholder is allocated a share generated from a private value ($i.e.$, the secret in classical secret sharing). Then, the dealer prepares three qudits and embeds two secrets and a verification value into each qudit respectively, moreover, each qudit also includes the same private value. These qudits are delivered along a line of at least $t$ participants. On receiving the qudits, each participant performs unitary operations related to the share on the qudits. On one hand, an operation add a random number to each secret and the verification value, on the other hand,  the private value in each qudit is eliminated due to classical ($t,n$)-SS after at least $t$ participants complete their unitary operations. Subsequently, the last participant measures the three qudits and publishes the measurement results to all participants. Finally, all participants recover the two secrets and the verification value after disclosing their respective random numbers. Each participant is able to check the correctness of the two secrets by the verification value, and thus can detect any cheats and eavesdropping during secret reconstruction.

\subsubsection{Proposed scheme}
\par The scheme consists of two phases, \emph{Classical private share distribution} and \emph{Secret sharing}, and we present each phase in detail as follows.
\par \emph{Classical private share distribution phase.} In this phase, dealer Alice distributes classical private shares to $n$ shareholders Bob$_j, j=1,2,...,n$.
\begin{enumerate}[($i$)]
	\item Alice picks a random polynomial $f(x)$ of degree at most $(t-1)$ over finite field GF$(d)$:
	\[f(x) = {a_0} + {a_1}x + ... + {a_{t - 1}}{x^{t - 1}}\bmod d,\]
	where $s = {a_0} = f(0)$ is the private value and all coefficients ${a_j}, j = 0,1,...,t - 1$, are in the finite field GF$(d)$ for a large prime $d$.
	\item Alice computes $f({x_j})$ as the share of shareholder Bob$_j$ for $j = 1,2,...,n$, where ${x_j} \in {\rm{GF}}(d)$ is the public information of Bob$_j$ with ${x_j} \ne {x_r}$ for $j \ne r$. $n,  (n\ge t),$ is the total number of shareholders. A shareholder is also called participant when it participates in secret reconstruction.
	\item Alice sends each share $f(x_j)$ to the corresponding shareholder Bob$_j$ through private channel, which guarantees shares are delivered securely from Alice to shareholders.
\end{enumerate}

\emph{Secret sharing phase.} The dealer Alice first prepares three identical states $\left| {{\Phi_v}} \right\rangle  = \left| {{\varphi_0}^0} \right\rangle  = \frac{1}{{\sqrt d }}\sum_{j = 0}^{d - 1} {\left| j \right\rangle },v = 1,2,3$, and then shares secrets $S_1,S_2\in {\rm{GF}}(d)$ and a check value $N\in {\rm{GF}}(d)$ among $m$, $(m\geq t)$ shareholders by taking the following steps.
\begin{enumerate}[($i$)]
	\item Alice performs the operations ${U_{{p_0}^v{q_0}^v}} = {X_d}^{{p_0}^v}{Y_d}^{{q_0}^v}$ on $\left| {{\Phi _v}} \right\rangle ,$ which transform the states $\left| {{\Phi _v}} \right\rangle $ into ${\left| {{\Phi _v}} \right\rangle _0} = \left| {{\varphi _{{p_0}^v}}^{{q_0}^v}} \right\rangle$, where $  {p_0}^1 = {S_1},{p_0}^2 = {S_2}$, ${p_0}^3 = N$, ${q_0}^1 = {q_0}^2 = {q_0}^3 = d - s$ with ${p_0}^v,{q_0}^v \in {\rm{GF}}(d) $, ${S_1} = {S_2}N\bmod d$.
	\item \label{II} Suppose that Alice needs to share secrets $S_1,S_2$ among $m$ $(m\geq t)$ participants $\{{\rm{Bo}}{{\rm{b}}_j},j=1,2,...,m\}$,  she sends the three states ${\left| {{\Phi _v}} \right\rangle _0}$ to Bob$_1$. Upon receiving the three states, Bob$_1$ performs operations ${U_{{p_1}^v{q_1}^v}}$ on ${\left| {{\Phi _v}} \right\rangle _0}$ respectively, where ${p_1}^v$ are mutually independent random numbers, ${q_1}^v = {c_1} = f({x_1})\prod\nolimits_{r = 2}^m {\frac{{{x_r}}}{{{x_r} - {x_1}}}} \bmod d, v=1,2,3$, and ${p_1}^v,{q_1}^v \in {\rm{GF}}(d) $. As a result, the states ${\left| {{\Phi _v}} \right\rangle _0}$ are transformed into ${\left| {{\Phi _v}} \right\rangle _1} = \left| {{\varphi _{({p_0}^v + {p_1}^v)}}^{({q_0}^v + {q_1}^v)}} \right\rangle $. Bob$_1$ delivers the states ${\left| {{\Phi _v}} \right\rangle _1}$ to Bob$_2$.
	\item Each of the other participants Bob$_j,j = 2,3,...,m,$ repeats the same procedure sequentially as Bob$_1$ does in ($\ref{II}$). That is, Bob$_j$ performs the operations ${U_{{p_j}^v{q_j}^v}}$ on ${\left| {{\Phi _v}} \right\rangle _{j - 1}}$ accordingly and thus gets the states ${\left| {{\Phi _v}} \right\rangle _j}= \left| {{\varphi _{\sum\nolimits_{r = 0}^j {{p_r}^v} }}^{\sum\nolimits_{r = 0}^j {{q_r}^v} }} \right\rangle $, where $ {p_j}^v,{q_j}^v \in {\rm{GF}}(d) $ , ${p_j}^v$ are mutually independent random numbers, ${q_j}^v = {c_j} = f({x_j})\prod\nolimits_{r = 1,r \ne j}^m {\frac{{{x_r}}}{{{x_r} - {x_j}}}} \bmod d$. Subsequently,	Bob$_j$ send ${\left| {{\Phi _v}} \right\rangle _j}$ to next participant  Bob$_{j + 1},j=2,3,...,m-1$.
	\item Consequently, the last participant Bob$_m$ keeps the three states and chooses the basis ${\left\{ {\left| {{\varphi _l}^0} \right\rangle } \right\}_l}$ to measure these three states. The results are labeled ${R_1}$, ${R_2}$, ${R_3}$ respectively and then Bob$_m$ publishes the results.
	\item The measurement basis is ${\left\{ {\left| {{\varphi _l}^0} \right\rangle } \right\}_l}$ due to
    \begin{equation} \label{4}
    \begin{aligned}
	\sum\limits_{j = 0}^m {{q_j}}  = d - s + \sum\limits_{j = 1}^m {{c_j}} \bmod d = 0\bmod d.
	\end{aligned}
    \end{equation}
In this case, measurement results ${R_1}$, ${R_2}$, ${R_3}$ and the random numbers ${p_j}^1,{p_j}^2,{p_j}^3,j = 0,1,...,m$ satisfy
	\begin{equation}
    \begin{aligned}
	\sum\limits_{j = 0}^m {{p_j}^v}  = {R_v}\bmod d,v = 1,2,3.
	\end{aligned}
    \end{equation}
	After all $m$ participants exchange their  random numbers, they obtain the values ${p_0}^1=R_1-\sum_{j = 1}^m {{p_j}^1}$, ${p_0}^2=R_2-\sum_{j = 1}^m {{p_j}^2}$ and ${p_0}^3=R_3-\sum_{j = 1}^m {{p_j}^3}$ respectively.
	\item \label{VI} To check the correctness of the values ${p_0}^1$, ${p_0}^2$ and ${p_0}^3$, each participant can verify whether the following equation holds
	\begin{equation}\label{6}{p_0}^1 = {p_0}^2{p_0}^3\bmod d.\end{equation} If it does, the secret sharing attempt is not corrupt and thus all participants share the dealer's secrets ${S_1}={p_0}^1$, ${S_2}={p_0}^2$; otherwise they are aware that the secret sharing is invalid and abort this round.
\end{enumerate}

\subsubsection{Correctness of the scheme}

The scheme is correct because, after the dealer and all $m$ $(m\geq t)$ participants complete their operations, each final state becomes

\begin{equation}
\begin{aligned}
{\left| \Phi  \right\rangle _m}
&=\left( {\prod\limits_{r = 0}^m {{X_d}^{{p_r}^v}{Y_d}^{{q_r}^v}} } \right)\left| \Phi  \right\rangle   \\
&=\frac{1}{{\sqrt d }}\sum\limits_{j = 0}^{d - 1} {{\omega ^{\sum\nolimits_{r = 0}^m {\left( {j{p_r}^v + {j^2}{q_r}^v} \right)} }}\left| j \right\rangle },
\end{aligned}
\end{equation}
for $v=1,2,3$.

Besides, all participants in Shamir's (\emph{t}, \emph{n})-SS can recover the secret by summing up all components. $i.e.$,
\begin{equation} \label{8}
\begin{aligned}
s = f(0)=\sum_{j=1}^{m}{c_j} \bmod d=\sum_{j=1}^{m}{f({x_j})\prod\limits_{r = 1,r \ne j}^m {\frac{{{x_r}}}{{{x_r} - {x_j}}}} \bmod d},
\end{aligned}
\end{equation}
which can be immediately obtained by Lagrange interpolation formula.

Thus we have $\sum_{j = 1}^m {{c_j}}=Nd+s,N \in Z$, so the Eq.(\ref{4}) holds.
Then due to Eq.(\ref{4}) which is $\sum\nolimits_{j = 0}^m {{q_j} = 0\bmod d} $, it follows that when Bob$_m$ measures the final states in the basis ${\left\{ {\left| {{\varphi _l}^0} \right\rangle } \right\}_l}$, he obtains the real results satisfying Eq.(\ref{6}). Finally, with published results, all $m$ participants can recover secrets by exchanging their random numbers. Here, the proposed scheme satisfies reliability, the first criteria of a well-defined ($t, n$)-SS since it allows at least $t$ participants recovering the secret.

\section{Security analysis}
In this section, we show that our scheme is perfect and secure against various attack strategies including intercept-resend attack, participant attack and joint attack. Then we will analyze the validity of the verification mechanism against cheat and eavesdropping.

\subsection{Confidentiality analysis}

Here, we mainly prove the perfect security of the scheme. A $(t,n)$-QSS scheme is perfect with respect to the probability distribution of secret over secret space if less than $t$ participants obtains no information about the secret.

\emph{Theorem 1}. The proposed $(t,n)$-QSS scheme is perfect with respect to the probability distribution of secret over secret space. That is,
\begin{equation}
I(S_v;{\Omega}) = H(S_v) - H(S_v|{\Omega}) =0
\end{equation}
where ${\Omega}$ denotes the set of shares available  for less than $t-1$ participants, $H(S_v)$ is the information entropy of the secret $S_v,v=1,2$ and $I(S_v;{\Omega})$ represents the mutual information of $S_v$ with ${\Omega}$.\\

\begin{proof}
Suppose that $m(m\ge{t}) $ participants $\{{\rm{Bo}}{{\rm{b}}_j}, j=1,2,...,m\}$, with the corresponding shares $\{f(x_j),j=1,2,...,m\}$, collaborate to recover each secret $S_v,v=1,2$ in normal case.

Without losing generality, assume exactly $t-1$ participants $\{{\rm{Bo}}{{\rm{b}}_j}, j=1,2,...,t-1\}$ conspire in the attack and Bob$_{t-1}$ measures ${\left| {{\Phi _v}} \right\rangle _{t-1}}$. But only with true measurement basis which is no longer the appointed basis ${\left\{ {\left| {{\varphi _l}^0} \right\rangle } \right\}_l}$, they can get correct results; otherwise due to Eq.(\ref{2}), they will get the correct results with the probability $1/d$. The key point is that the measurement basis is encoded by the private value $s$ used in \emph{Classical private share distribution phase}. But according to Shamir's\cite{Shamir1979} $(t,n)$-SS, for $m\ge{t}$, each private value
\begin{equation}
s=\sum_{j=1}^{m}{f({x_j})\prod\limits_{r = 1,r \ne j}^m {\frac{{{x_r}}}{{{x_r} - {x_j}}}} \bmod d},
\end{equation}
and $s$ is uniformly distributed over GF$(d)$ if $t-1$ participants $\{{\rm{Bo}}{{\rm{b}}_j}, j=1,2,...,t-1\}$ conspire. That is, with the shares $\Omega=\{f(x_j),j=1,2,...,t-1\}$ available, they have the probability $P(s|\Omega)=1/d$ to obtain $s$. Further, they can get correct results and random numbers exchanged from others, then recover the secrets. Thus the probability to obtain $S_v$ is $1/d$. As a result, we have conditional entropy $H(S_v|\Omega)=\log d$.

On the other hand, since ${p_{v0}}$ is uniformly and randomly selected from GF$(d)$ by Alice from the view of participants, hence, $S_v={p_{v0}}$ is indistinguishable from a random variable uniformly distributed over GF$(d)$, $i.e.$, $P(S_v)=1/d$. Consequently, the entropy of $S_v$ is $H(S_v)=\log d$.

Therefore, we finally have
\begin{equation}
I(S_v;{\Omega}) = H(S_v) - H(S_v|{\Omega}) =0.
\end{equation}

Since both secrets $S_v,v=1,2$ and the verification value $S_3$ all satisfy the above equation, the proposed scheme is perfect with respect to probability distribution of secrets in the secret space GF$(d)$. Here, the criteria confidentiality is satisfied.
\end{proof}

Overall, we can see that with respecting the two criteria, the proposed scheme is a well-defined ($t, n$) QSS.

\subsection{Intercept-resend attack}
Here, we consider the intercept-resend attack mounted by an external eavesdropper Eve. She may intercept the qudit $\left| {{\varphi _l}^k} \right\rangle $ sent from Bob$_j$ to Bob$_{j+1}$, but with no information about the measurement basis. Obviously, Eve can obtain the correct measurement result only when she happens to choose the true basis $k' = k$, which has the probability of ${1 \mathord{\left/{\vphantom {1 d}} \right.\kern-\nulldelimiterspace} d}$. Moreover, the correct measurement result is the sum of dealer's secret and preceding participants' random numbers, she can infer the dealer's secret only with the probability ${1 \mathord{\left/{\vphantom {1 d}} \right.\kern-\nulldelimiterspace} d}$  if she doesn't know these random numbers. Relatively, she will fail and change the qudit sent to Bob$_{j + 1}$ with probability ${{(d - 1)} \mathord{\left/{\vphantom {{(d - 1)} d}} \right.\kern-\nulldelimiterspace} d}$. Then it will cause contradiction to Eq.(\ref{6}) and thus be detected in step ($\ref{VI}$) in the scheme. In a word, Eve cannot figure out the secret with the probability more than ${1 \mathord{\left/{\vphantom {1 d}} \right.\kern-\nulldelimiterspace} d}$ in intercept-resend attack. Obviously, the scheme is more secure for a larger prime $d$.

\subsection{Participant attack}
A simple attack strategy a participant may take is using a random number, instead of the component $c_j$ generated by his own share, in the unitary operation. However, this attack will not be effective since the Eq.(\ref{4}) is violated and thus the last participant Bob$_m$ cannot obtain correct measurement results with the basis ${\left\{ {\left| {{\varphi _l}^0} \right\rangle } \right\}_l}$. That is, the published measurement results are random numbers in GF$(d)$. Therefore, after participants exchange random numbers, they all get wrong secrets. Consequently, this attack will be detected in step ($\ref{VI}$) because of the violation of Eq.(\ref{6}).

Considering that the first participant Bob$_1$ tries to infer the dealer's secrets alone by measuring qudits sent directly from the dealer. In this case, with only one share ($i.e.$, less than $t$ shares), he cannot recover the private value \emph{s} previously embedded in qudits by the dealer, because Shamir's ($t, n$)-SS is perfect \cite{Miao2015}, $i.e.$, no information about the secret can be obtained in Shamir's ($t, n$)-SS with less than $t$ shares. Thus, he can never measure the qudits in true basis but only to guess with the neglible probability $1/d$. It means this attack works to get secret with probability $1/d$ which is the same to guess the secret.

The last participant is at the special role in the proposed scheme, with the opportunity to measure the qudits in true basis. Thus in section \ref{Verification mechaism}, we will analyze the attack mounted by the last participant, which is more challenging to the scheme's security.
\subsection{Joint attack}

Considering the joint attack taken by part of participants in association with entanglement swapping \cite{Gao2007,He2007}, they could entangle the qudit with an ancilla ($e.g.$ Bell states), or use new entangled states to replace the qudits. However, they benefit nothing from this attack, because less than $t$ shares cannot recover the private value $s$, even though entanglement swapping renders the qudits available for cheaters in a mixed state, there is no observable result obtained. As a result, they will face the problems that possessing no knowledge on measurement basis and also that they cannot achieve random numbers used by other honest participants. Furthermore, this attack can be detected in step ($\ref{VI}$) because the basis ${\left\{ {\left| {{\varphi _l}^0} \right\rangle } \right\}_l}$ is wrong and consequently the recovered values do not satisfy ${p_0}^1 = {p_0}^2{p_0}^3\bmod d$.

\subsection{Verification mechanism}
\label{Verification mechaism}
The last participant Bob$_m$ is crucial to our scheme because he is responsible for keeping and measuring the qudits in true basis. So, he is able to deceive the other participants by announcing wrong measurement results. Of course, other participant can also cheat by using a wrong share in secret reconstruction. Moreover, the qudits are obviously vulnerable to eavesdropping. Thus it is necessary to establish appropriate verification mechanism to detect cheat or eavesdropping. As mentioned above, the proposed scheme uses Eq.(\ref{6}) as the verification mechanism.

Let consider the error rate of verification mechanism, $i.e.$, the probability that the verification mechanism does not detect wrong secrets.

For correct measurement results ${R_v}$ and  the corresponding sums of random numbers of all participants, ${N_v=\sum_{j=1}^{m}{p_j}^v} , v=1,2,3$, assume that the last participant  Bob$_m$ publishes wrong measurements ${R_v}^\prime \neq{R_v} , v=1,2,3$. Obviously, if $({R_1}^\prime-{N_1})=({R_2}^\prime-{N_2})({R_3}^\prime-{N_3})\bmod d$ happens to hold, the wrong measurement results cannot be detected. In this case, the verification mechanism fails and thus the other participants recover wrong secrets without being detected.

Obviously, if Bob$_m$ randomly and uniformly chooses three values ${R_1}^\prime, {R_2}^\prime$ and ${R_3}^\prime$ in GF$(d)$ as measurement results and publishes them to the other participants, there are totally $d^3$ tuples of  $\{{R_1}^\prime, {R_2}^\prime,{R_3}^\prime\}$. Note that  ${R_v}$ are published before all participants exchange their random values ${p_j}^v,j=1,2,...,m,$ to obtain the sums ${N_v} $. Since each participant Bob$_j$ picks its random values ${p_j}^v$ privately and independently, ${N_v=\sum_{j=1}^{m}{p_j}^v} $ , are indistinguishable from random numbers uniformly distributed in GF$(d)$ in the view of all participants. As a result, $({R_v}^\prime-{N_v})$ are also indistinguishable from random numbers uniformly distributed in GF$(d)$ for participants. In this case, there are totally $d^2$ randomly selected tuples of $\{{R_1}^\prime, {R_2}^\prime,{R_3}^\prime\}$ satisfying
\begin{equation}
\begin{aligned}
  ({R_1}^\prime-{N_1})=({R_2}^\prime-{N_2})({R_3}^\prime-{N_3})\bmod  d,
\end{aligned}
\end{equation}
because, given $\{N_1,N_2,N_3\},$ ${R_3}^\prime$ can always be determined for randomly selected pairs of $\{{R_1}^\prime, {R_2}^\prime\}$.

The result is the same if any other participant cheats by using a wrong share when performing operations on quantum states.

Therefore, the error rate of the verification mechanism is $d^2/d^3=1/d$.
Since $d$ is a large prime, the error rate converges to 0 when $d$ approaches to infinity.

In conclusion, the scheme can detect the cheat by participants with the probability $(d-1)/d$, which converges to $100\%$ if $d$ approaches to infinity.

\section{Comparisons and discussion}

There are many QSS schemes, but most of them are 2-level \cite{Hillery1999,Guo2003,Tokunaga2005,Qin2015,Hsu2003,Schmid2005,Markham2008} and with ($n$, $n$) structure \cite{Hillery1999,Guo2003,Yu2008,Bai2017,Tavakoli2015,Hsu2003,Schmid2005,Markham2008,Karimipour2015,Yang2013,Lai2016_2,Kogias2017}. For instance, in the scheme \cite{Schmid2005}, the authors use phase shift operation to embed the secret into a single qubit such that the secret can be recovered after all participants complete their operations. Besides, a special QSS based on Grover quantum searching algorithm was proposed in \cite{Hsu2003}. But these 2-level schemes have less universality and practicability when compared to $d$-level ones and ($n$, $n$) structure QSS schemes are less flexible than ($t$, $n$) ones in the sense that, other than any $t$ parties, all shareholders must be present to recover the secret. Compared with these schemes, our $d$-level ($t,n$) threshold quantum secret sharing scheme is more flexible, universal and practicable. Hence, the following parts concern about $d$-level or ($t$, $n$) threshold structure schemes.

The scheme in \cite{Yu2008} initially prepares a high-fidelity entangled GHZ state with $n$ subsystems. Once the state is produced, the number of participants is fixed. Yu $et$ $al.$ presented another QSS \cite{Bai2017} based on $d$-dimensional GHZ state, which is also not scalable with the growth of participant number. In their scheme, an X-basis measurement and classical communication are used to distinguish two orthogonal states and reconstruct the original secret. Some $d$-level schemes \cite{Yang2013,Song2017} were proposed based on Quantum Fourier Transform. The scheme in \cite{Yang2013} disguises each share of a secret with true randomness, rather than classical pseudo randomness. But schemes using entangled states will not be scalable with growing participants. Also a common problem of these schemes is that each participant needs to measure his particle at last, but some participant may fail in measurement due to inefficient detection and thus render an invalid secret sharing easily. Compared with these schemes based on special quantum system, our scheme enjoys a strong scalability because it is almost not restricted by participant number in realization.
\par QSS schemes with ($t$, $n$) structure was first proposed in 1999 \cite{Cleve1999} based on quantum error correcting code. The scheme divides a special quantum state into $n$ shares, such that any $t$ or more than $t$ participants can recover the initial state using linear transformation. However, it is hard to map the quantum state to $n$ quantum states in coding. Later, some other threshold schemes are proposed with different physical characteristics, such as those in \cite{Lance2003,Lau2013,Wu2014} benefit from continuous variable and in \cite{Rahaman2015,Yang2015,Wang2017} construct from the ability of exactly distinguishing orthogonal multipartite entangled states under restricted local operation and classical communication. Some schemes \cite{Tokunaga2005,Qin2015,Lai2016,Song2017,Lu2018} take advantage of the classical ($t, n$)-SS, which using phase shift operation to embed the secret and shares generated from classical ($t, n$)-SS into processed quantum state, so after sequential operations, participants can collaborate to recover secret.

Compared with these previous schemes in Table \ref{table 1}, our scheme employs $d$-level unitary operation in association with classical ($t,n$)-SS. Due to the verification mechanism, it is free from the trusted third party who is responsible for measurement results and any cheat behavior of a participant can be detected easily.

\begin{table}[]
	\centering
\caption{Comparisons with previous QSSs.}\label{table 1}	
	\begin{tabular}{|l|c|c|c|c|c|}
		\hline
        \rowcolor{mygray} Schemes & Ref.\cite{Shamir1979} &Ref.\cite{Cleve1999}&Ref.\cite{Hillery1999}&Ref.\cite{Tokunaga2005}&Our scheme\\ \hline Entanglement-free &  &No&No&Yes&Yes\\ \hline
        Scalability & Yes &No&No&Yes&Yes\\ \hline
        Level &  &$d$&2&2&$d$\\ \hline
        ($t, n$) threshold & Yes &Yes&No&Yes&Yes\\ \hline
        Cheat detection & No &No&No&No&Yes\\ \hline
	\end{tabular}

\end{table}

Thinking further about the proposed scheme, we can find a generic method to construct such type of $d$-level threshold QSS schemes. Note that our scheme employs the classical Shamir's ($t,n$) secret sharing, in fact, other classical ($t$, $n$)-secret sharing schemes, such as linear code based ($t$, $n$)-SS \cite{McEliece1981,Massey1993}, geometry based ($t$, $n$)-SS \cite{Blakley1979}, Chinese Remainder Theorem based ($t$, $n$)-SS \cite{Asmuth1983,Mignotte1983}, etc., can also be directly used to construct new threshold QSS schemes based on a single $d$-level quantum system. All these new schemes share the same features as our proposed scheme. Furthermore, each participant, $e.g.$, Bob$_j$, constructs a component c$_j$ from the share and then produces the $d$-level unitary operations from c$_j$. After each participant completing its $d$-level unitary operation on a qudit sequentially, all components c$_j,j=1,2,...m$, are actually added up and the private value $s$ is removed, which ensures that the last participant Bob$_m$  gets the correct measurement result. As a matter of fact, as long as a classical ($t,n$)-SS has the property of cumulative sum, $i.e.$, the secret ($i.e.$, private value in our scheme) \emph{s}, can be  expressed as $s = \sum\nolimits_{j = 1}^m {{c_j}} \bmod M = \sum\nolimits_{j = 1}^m {{a_j}} {s_j}\bmod M$, it can be used to construct such a $d$-level  ($t,n$) threshold quantum secret sharing , where ${c_j}$ are the values Bob$_j$ evaluated from the shares ${s_j}$ and ${a_j}$ are some public parameters, $m \ge t$ is the number of participants and $M$ is a modulus.

\section{Conclusion}
This paper presents a ($t,n$) threshold QSS scheme based on a single $d$-level quantum system. The scheme simply requires sequential communication of a single $d$-level quantum system during secret reconstruction. It is flexible in application, scalable to participant number and easy to realize. Security analyses show the scheme is secure against typical attacks. Moreover, a verification mechanism used to verify recovered secrets so that eavesdropping and cheats can be detected.

By the method of our scheme, new ($t,n$) threshold QSS schemes based on a single $d$-level quantum system can be easily constructed if the Shamir's ($t,n$) secret sharing scheme is replaced by other classical threshold ones.

\section*{Acknowledgement}
We would like to thank the anonymous reviewer for helpful suggestions. This work is supported by the National Natural Science Foundation of China under 61572454, 61572453, 61520106007.


\begin{thebibliography}{99}
\bibitem{Shamir1979} Shamir, A.: How to share a secret. Commun. ACM 22, 612 (1979)
\bibitem{Blakley1979} Blakley, G.R.: Safeguarding cryptographic keys. Proc. Nat. Comput. Conf., New York, 313 (1979)
\bibitem{Harn2013} Harn, L.: Group authentication. IEEE Trans. Comput. 62(9), 1893 (2013)
\bibitem{Boldyreva2002} Boldyreva, A.: Threshold signatures, multisignatures and blind signatures based on the gap-Diffie-Hellman-group signature scheme. Public Key cryptography-PKC., Berlin, Germany: Springer-Verlag, 31 (2002)
\bibitem{Harn1994} Harn, L.: Group-oriented ($t$, $n$) threshold digital signature scheme and digital multisignature. IEE Proc. Comput. Digit. Techn. 141(5), 307 (1994)
\bibitem{Liu2016} Liu, Y.N., Harn, L., Mao, L., Xiong, Z.: Full-healing group-key distribution in online social networks. International Journal of Security and Networks 11(1-2), 12 (2016)
\bibitem{Desmedt1994} Desmedt, Y.G.: Threshold cryptography. Eur. Trans. Telecommun. 5(4), 449 (1994)
\bibitem{Patel2016} Patel, K.: Secure multiparty computation using secret sharing. International Conference on Signal Processing, Communication, Power and Embedded System, IEEE, p. 863 (2016)


\bibitem{Cleve1999} Cleve, R., Gottesman, D., Lo, H.K.: How to share a quantum secret. Phys. Rev. Lett 83, 648 (1999)
\bibitem{Hillery1999} Hillery, M., Buzek, V., Berthiaume, A.: Quantum secret sharing. Phys. Rev. A 59, 1829 (1999)

\bibitem{Yu2008} Yu, I.C., Lin, F.L., Huang, C.Y.: Quantum secret sharing with multilevel mutually (un) biased bases. Phys. Rev. A 78, 012344 (2008)
\bibitem{Bai2017} Bai, C.M., Li, Z.H., Xu, T.T., Li, Y.M.: Quantum secret sharing using the $d$-dimensional GHZ state. Quantum Inf. Process 16(3) (2017)
\bibitem{Tavakoli2015} Tavakoli, A., Herbauts, I., Zukowski, M., Bourennane, M.: Secret sharing with a single $d$-level quantum system Phys. Rev. A 92, 030302 (2015)
\bibitem{Hsu2003} Hsu, L.Y.: Quantum secret-sharing protocol based on Grover's algorithm. Phys. Rev. A 68, 022306 (2003)
\bibitem{Guo2003} Guo, G.P., Guo, G.C.: Quantum secret sharing without entanglement. Phys. Lett. A 310(4), 247-251 (2003)
\bibitem{Markham2008} Markham, D., Sanders, B.C.: Graph states for quantum secret sharing. Phys. Rev. A 78, 042309 (2008)
\bibitem{Karimipour2015} Karimipour, V., Asoudeh, M.: Quantum Secret Sharing and Random Hopping: Using single states instead of entanglement. Phys. Rev. A 92, 030301 (2015)
\bibitem{Yang2013} Yang, W., Huang, L., Shi, R., He, L.: Secret sharing based on quantum Fourier transform. Quantum Inf. Process 12(7), 2465 (2013)
\bibitem{Lai2016_2} Lai, H., Luo, M.X., Pieprzyk, J., Li, T., Liu, Z.M., Orgun, M.A.: Large-capacity three-party quantum digital secret sharing using three particular matrices coding, Commun. Theor. Phys. 66(05), 501-508 (2016)
\bibitem{Kogias2017} Kogias, I., Xiang, Y., He, Q.Y., Adesso, G.: Unconditional security of entanglement-based continuous-variable quantum secret sharing. Phys. Rev. A 95, 012315 (2017)


\bibitem{Tittel2001} Tittel, W., Zbinden, H., Gisin, N.: Experimental demonstration of quantum secret sharing. Phys. Rev. A 63(4), 042301 (2001)
\bibitem{Schmid2005} Schmid, C., Trojek, P., Bourennane, M., Kurtsiefer, C., Zukowski, M., Weinfurter, H.: Experimental single qubit quantum secret sharing. Phys. Rev. Lett. 95, 230505 (2005)
\bibitem{Chen2005} Chen, Y.A., Zhang, A.N., Zhao, Z., Zhou, X.Q., Lu, C.Y., Peng, C.Z., $et$ $al.$: Experimental quantum secret sharing and third-man quantum cryptography. Phys. Rev. Lett 95(20), 200502 (2005)

\bibitem{Lu2016} Lu, H., Zhang, Z., Chen, L.K., Li, Z.D., Liu, C., Li, L., $et$ $al.$: Secret sharing of a quantum state. Phys. Rev. Lett 117(3), 030501 (2016).

\bibitem{Lance2003} Lance, A.M., Symul, T., Bowen, W.P., Tyc, T., Sanders, B.C., Lam, P.K.: Continuous variable (2, 3) threshold quantum secret sharing schemes. New J. Phys. 5, 4 (2003)
\bibitem{Lau2013} Lau, H.K., Weedbrook, C.: Quantum secret sharing with continuous-variable cluster states. Phys. Rev. A 88, 042313 (2013)
\bibitem{Wu2014} Wu, Y., Cai, R., He, G., Zhang, J.: Quantum secret sharing with continuous variable graph state. Quant. Inf. Process. 13, 1085 (2014)
\bibitem{Rahaman2015} Rahaman, R., Parker, M.G.: Quantum scheme for secret sharing based on local distinguishability. Phys. Rev. A 91, 022330 (2015)
\bibitem{Yang2015} Yang, Y.H., Gao, F., Wu, X., Qin, S.J., Zuo, H.J., Wen, Q.Y.: Quantum secret sharing via local operations and classical communication. Sci. Rep. 5, 16967 (2015)
\bibitem{Wang2017} Wang, J., Li, L., Peng, H., Yang, Y.: Quantum-secret-sharing scheme based on local distinguishability of orthogonal multiqudit entangled states. Phys. Rev. A 95, 022320 (2017)
\bibitem{Tokunaga2005} Tokunaga, Y., Okamoto, T., Imoto, N.: Threshold quantum cryptography. Phys. Rev. A 71, 012314 (2005)
\bibitem{Qin2015} Qin, H., Zhu, X., Dai, Y.: ($t$, $n$) Threshold quantum secret sharing using the phase shift operation. Quantum Inf. Process 14(8) (2015)

\bibitem{Song2017} Song, X.L., Liu, Y.B., Deng, H.Y., Xiao, Y.G.: ($t$, $n$) Threshold d-level Quantum Secret Sharing. Sci. Rep. 7, 6366 (2017)
\bibitem{Lu2018} Lu, C.B., Miao, F.Y., Meng, K.J., Yu, Y.: Threshold quantum secret sharing based on single qubit. Quantum Inf. Process, 17(3):64 (2018)
\bibitem{Lai2016} Lai, H., Zhang, J., Luo, M.X., Pan, L., Pieprzyk, J., Xiao, F.Y., Orgun, M.A.: Hybrid threshold adaptable quantum secret sharing scheme with reverse Huffman-Fibonacci tree coding, Sci. Rep. 6, 31350 (2016)
\bibitem{Ivanovic1981} Ivanovic, I.D.: Geometrical description of quantal state determination. J. Phys. A 14, 3241 (1981)
\bibitem{Wootters1989} Wootters, W.L., Fields, B.D.: Optimal state-determination by mutually unbiased measurements. Ann. Phys. 191, 363 (1989)

\bibitem{Miao2015} Miao, F.Y., Xiong, Y., Wang, X.F., Badawy, M.: Randomized Component and Its Application to
(t,m,n)-Group Oriented Secret Sharing. IEEE. T. INF. FOREN. SEC, 10(5) (2015)
\bibitem{Gao2007} Gao, F., Qin, S.J., Wen, Q.Y.: A simple participant attack on the Bradler-Dusek protocol. Quantum Inf. Comput. 7(4), 329 (2007)
\bibitem{He2007} He, G.P.: Comment on "experimental single qubit quantum secret sharing". Phys. Rev. Lett. 98, 028901 (2007)
\bibitem{McEliece1981} McEliece, R.J., Sarwate, D.V.: On sharing secrets and Reed-Solomon codes. Commun. ACM 24(9), 583 (1981)
\bibitem{Massey1993} Massey, J.L.: Minimal codewords and secret sharing. Proceedings of the 6th Joint Swedish-Russian International Workshop on Information Theory. Washington D C, IEEE Press, 276 (1993)
\bibitem{Asmuth1983} Asmuth, C., Bloom, J.: A modular approach to key safeguarding. IEEE. T. Inform. Theory 30(2), 208 (1983)
\bibitem{Mignotte1983} Mignotte, M.: How to Share a Secret. Conference on Cryptography, Berlin, Germany: Springer-Verlag 149, 371 (1982)
\end{thebibliography}
\end{document}